\documentclass[runningheads]{llncs}
\usepackage[T1]{fontenc}
\usepackage{graphicx}
\usepackage{multirow}
\usepackage{caption}
\usepackage[table,xcdraw,svgnames]{xcolor}
\usepackage{subfigure}
\usepackage{tikz,xcolor,hyperref}
\usepackage{amsmath}
\usepackage{amsfonts}
\usepackage{pifont}
\DeclareMathOperator*{\argmin}{arg\,min}
\definecolor{lime}{HTML}{A6CE39}
\DeclareRobustCommand{\orcidicon}{
	\begin{tikzpicture}
	\draw[lime, fill=lime] (0,0) 
	circle [radius=0.16] 
	node[white] {{\fontfamily{qag}\selectfont \tiny ID}};
	\draw[white, fill=white] (-0.0625,0.095) 
	circle [radius=0.007];
	\end{tikzpicture}
	\hspace{-2mm}
}
\foreach \x in {A, ..., Z}{\expandafter\xdef\csname orcid\x\endcsname{\noexpand\href{https://orcid.org/\csname orcidauthor\x\endcsname}
			{\noexpand\orcidicon}}
}

\begin{document}
\title{An Empirical Study of the Imbalance Issue in Software Vulnerability Detection\thanks{This work is funded by the European Union's Horizon Research and Innovation Programme under Grant Agreement n$^\circ$101070303.}}
\titlerunning{Imbalance in Software Vulnerability Detection}

\author{Yuejun Guo\inst{1}\orcidA{} \and
Qiang Hu\inst{2}\orcidB{}\thanks{Corresponding author.} \and
Qiang Tang\inst{1} \orcidC{} \and
Yves Le Traon\inst{2}\orcidD{}}
\authorrunning{Y. Guo et al.}
\institute{ITIS, Luxembourg Institute of Science and Technology, Luxembourg \\
\email{\{yuejun.guo,qiang.tang\}@list.lu}
\and
SnT, University of Luxembourg, Luxembourg \\
\email{\{qiang.hu,yves.letraon\}@uni.lu}}

\maketitle 

\begin{abstract}
Vulnerability detection is crucial to protect software security. Nowadays, deep learning (DL) is the most promising technique to automate this detection task, leveraging its superior ability to extract patterns and representations within extensive code volumes. Despite its promise, DL-based vulnerability detection remains in its early stages, with model performance exhibiting variability across datasets. Drawing insights from other well-explored application areas like computer vision, we conjecture that the imbalance issue (the number of vulnerable code is extremely small) is at the core of the phenomenon. To validate this, we conduct a comprehensive empirical study involving nine open-source datasets and two state-of-the-art DL models. The results confirm our conjecture. We also obtain insightful findings on how existing imbalance solutions perform in vulnerability detection. It turns out that these solutions perform differently as well across datasets and evaluation metrics. Specifically: 1) \emph{Focal loss} is more suitable to improve the precision, 2) \emph{mean false error} and \emph{class-balanced loss} encourages the recall, and 3) \emph{random over-sampling} facilitates the F1-measure. However, none of them excels across all metrics. To delve deeper, we explore external influences on these solutions and offer insights for developing new solutions.

\end{abstract}

\section{Introduction}
\label{sec:intro}
The existence of software vulnerability is an inevitable risk in the software development life cycle, which raises significant concern since the vulnerability can be exploited by cybercriminals to run malicious code, install malware, and steal sensitive data. Discovering vulnerabilities in advance of the final deployment is ever required to enhance software security. 

\begin{figure}[ht]
    \centering
    \includegraphics[scale=0.85]{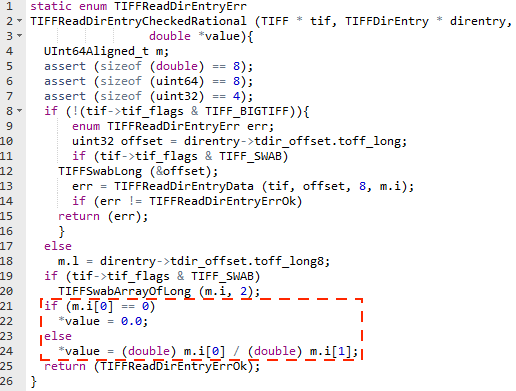}
    \caption{An example of vulnerable code: a C-language function from the LibTIFF project~\cite{libtiffweb}. This function is tagged with the ``Denial of Service'' vulnerability in the Lin2018 dataset (please refer to Section~\ref{subsec:data} for more details). The code framed in a red rectangle highlights a concern about handling cases of division by zero when $m.i[1]==0$.}
    \label{fig:vul_code}
\end{figure}

Manually identifying a function as vulnerable or not is tough concerning the required domain expertise and time. Fortunately, the rapid progress of deep learning (DL) largely automates this process~\cite{ml4code2021}. In this paper, we are interested in applying DL to software vulnerability detection at the function level, which enables early detection of vulnerabilities during the programming stage. Figure~\ref{fig:vul_code} shows an example of a vulnerable function tagged with ID CVE-2017-7597\footnote{\url{https://www.cvedetails.com/cve/CVE-2017-7597/?q=CVE-2017-7597}}. The detailed information is that tif\_dirread.c in LibTIFF 4.0.7 has an ``outside the range of representable values of type float'' undefined behavior issue, which might allow remote attackers to cause a denial of service (application crash) or possibly have unspecified other impacts via a crafted image. 

Although various DL models~\cite{saikat2022yet,rozo2021shallow,codebert2020,graphcodebert2020} have been designed for vulnerability detection, the imbalance issue that causes a high false positive rate or false negative rate~\cite{smote2002chawla} is usually ignored. In this paper, ``imbalance'' refers to the numeral difference where the number of vulnerable code is much less than secure code in a dataset. In practice, software vulnerabilities do exist but rarely (e.g., 1 vulnerability in every 500 C-language functions) in source code programmed by experienced software developers. This imbalance problem is widely studied in other areas, such as computer vision (CV) and natural language processing (NLP), and to tackle it, both data and model-level methods have been proposed. The main idea is to put more importance on the minority set. Data-level methods straightforwardly down-sample~\cite{Chris2003down} the majority set or over-sample~\cite{over2018yang} the minority set to ensure the numeral balance of data. Model-level approaches focus on the loss function~\cite{focal2020lin,cbl2019cui} that guides the training procedure via adjusting the weights of majority and minority sets.

Although DL-based vulnerability detection is gaining attention, how the imbalance issue affects the ``performance'' of DL models in this specific area is still an open question. What makes things more complex is that, the model performance is evaluated by various metrics in the literature~\cite{Li2019study}, such as accuracy~\cite{codexglue2021paper}, precision, and false positive rate. This makes it very difficult to compare their performance. For instance, the state-of-the-art (SOTA) model CodeBERT~\cite{codexglue2021paper} is reported to achieve over 60\% detection accuracy, while the false negative rate is 70\%. The model is acceptable if accuracy is the evaluation criterion but useless in the case of a false negative rate. 

In this paper, we conduct a comprehensive study on the imbalance issue in DL-based vulnerability detection. Through a series of experiments, our findings are summarized as follows. First of all, our first experimental results show that the imbalance problem tends to cause a model to gain a relatively low loss on secure code compared to the vulnerable one during the training procedure. Thus, The false negative rate is high.
    
Second, we experiment on how the imbalance solutions adapted from the CV and NLP domains perform in vulnerability detection. Our results show that:

-- Compared to accuracy and false positive rate, precision, recall, and F1-measure are more informative when evaluating a DL model for vulnerability detection.

-- None of the existing solutions from other domains performs perfectly across all selected datasets and models. Model-level solutions are beneficial to precision and recall. Data level solutions are more helpful to F1-measure.
    
Third, apart from the designed methodology, we explore how external factors affect the effectiveness of existing solutions, where these factors come from source code, such as the appearance of vulnerability types in the training procedure and test time and the detection difficulty of vulnerable code. Our experiment results show that external factors, such as the absence of vulnerabilities, identification difficulty of certain vulnerability types, and data distribution need to be considered when designing a new solution specifically for vulnerability detection.

For the readers to validate our findings, the experiment datasets and artifacts (including all solutions) for reproduction are made available on Git\footnote{\url{https://github.com/testing-cs/vulnerability-detection.git}}.

\section{Background and Related Work}
\label{sec:related}

\subsection{Software Vulnerability Detection}
\label{subsec:security}
A software vulnerability is a security flaw and glitch found in source code. Detecting vulnerabilities has attracted considerable interest in the security community. Manually checking source code is straightforward, but even for experts, this task is tedious and subjective because of great code complexity and diverse programming languages. At a higher level, there is the popular fuzzing~\cite{tradiontal2017vul} technique that automates the task. The basic idea of fuzzing is to generate a large number of test cases that are fed into the target program for execution. When a crash is triggered, it will be first determined as a bug or not and further identified as a vulnerability or not by exploitability analysis~\cite{fell2017review}. Since fuzzing highly relies on the generated test cases and the target program is required to be executed to monitor the behavior, it cannot be applied during a very early stage, such as the programming time. 

Traditional detection tools like static analyzers~\cite{Arusoaie2017comparison} often require manual feature engineering by security researchers and target specific vulnerabilities, which is less efficient~\cite{shen2020survey}. Deep learning (DL) facilitates the static analysis without running the target program and is feasible to function at different code module granularity~\cite{data2021vulreview}, such as file level~\cite{Aayush2022historical}, function level~\cite{devign2019zhou,codexglue2021paper,lin2018cross,deeplin2020}, and program slice level~\cite{vuldeepecker2018li,sysevr2022li}. Various DL models have been developed in the literature to support automated vulnerability detection. Simple examples~\cite{Li2019study} include multi-layer perception (MLP), convolutional neural network (CNN), long short-term memory (LSTM), gated recurrent unit (GRU), bidirectional LSTM, and bidirectional GRU. Advanced ones dedicated to the structural representation of source code with graph neural networks, such as Devign~\cite{devign2019zhou} proposed by Zhou \emph{et al.}. More recently, the application of foundation models is changing the domination of these task-specific models, which is discussed in the next subsection. 

\subsection{Recap of (foundation) DL Models}
\label{subsec:found}
In classical DL, a model is initially randomly parameterized. Given a large set of labeled data, the model is trained with a certain number of epochs to achieve a satisfying performance for a given task. Therefore, this type of model is also called the task-specific model. Depending on the data type, model architecture, and learning task, the training procedure can take minutes or days. For instance, given the ImageNet-1k dataset to obtain an image classification model with ResNet-50, the 90-epoch training takes 14 days on a NVIDIA M40 GPU~\cite{you2017imagenet}.

Different from task-specific models, foundation models, aka pre-trained models~\cite{han2021pre}\footnote{The term ``foundation model'' is used in this paper because, in the literature, a ``pre-trained model'' also has the meaning of a model trained by someone else and targeting a similar task\cite{surprise2019kim,choi2022pre}.}, break the limitation of relying on labeled data and have been a new paradigm of artificial intelligence (AI)~\cite{foundation}. Generally, a foundation model is trained using a huge volume of unlabeled data at scale and can be used for a wide range of downstream tasks. Via a few epochs' fine-tuning, the model can achieve SOTA performance. Foundation models have been increasingly developed and brought dramatic improvements in various communities, such as computer vision, natural language processing, and software engineering. Example models are the bidirectional encoder representations from transformers (BERT)~\cite{bert2019devlin}, generative pre-trained transformer 3 (GPT-3)~\cite{gpt32020tom}, Roberta~\cite{roberta2019}, ViLBERT~\cite{vilbert2019}, VideoBERT~\cite{videobert2019}, CodeBERT~\cite{codebert2020}, and GraphCodeBERT~\cite{graphcodebert2020}.

\subsection{Solutions for Addressing Imbalance}
\label{subsec:imb}
There are mainly two types of methods for the imbalance issue, data level and model level~\cite{BUDA2018249}. Data level solutions focus on balancing the data size between minority and majority, such as down-sampling on majority and over-sampling on minority. The basic re-sampling strategy is in a random manner where samples are randomly selected to be removed~\cite{Chris2003down} or duplicated~\cite{yin2021learning}. An advanced over-sampling is to introduce new data based on neighboring samples~\cite{smote2002chawla} or synthesized~\cite{over2018yang}. However, advanced solutions~\cite{Mendoza2022on} are data-dependent and some are inapplicable to source code. For instance, SMOTE~\cite{smote2002chawla} generates a new sample by joining $k$ minority class nearest neighbors in the feature space, but generated code are likely to be invalid concerning both the syntax and semantics. Shu \emph{et al.}~\cite{shu2022dazzle} proposed Dazzle that leverages the Wasserstein Generative Adversarial networks as an over-sampling solution for software vulnerability detection. However, how to ensure the correctness of generate code is not addressed. 

Model-level solutions, also known as cost-sensitive learning~\cite{cbl2019cui}, assume the costs (represented by the loss) caused by different errors are unequal. The most common way is to put higher weight on the loss of the minority and less on the majority. Typical solutions include calculating the loss on minority and majority separately~\cite{seperated2016wang}, effective number-based re-weighting~\cite{cbl2019cui}, and misclassification-focused~\cite{focal2020lin}. All these solutions adjust the decision boundary during the training procedure. Another method, threshold-moving~\cite{He2013thre} adjusts the boundary in the test time, which is simple but sensitive to the change in data. 

Most existing solutions are initially proposed and studied in computer vision~\cite{cbl2019cui,focal2020lin,BUDA2018249} and natural language processing~\cite{cbl2019cui,seperated2016wang}. In this paper, we investigate their effectiveness for software vulnerability detection. 

\section{Empirical Study Design}
\label{sec:study}
\begin{figure}[ht]
    \centering
    \includegraphics[scale=0.55]{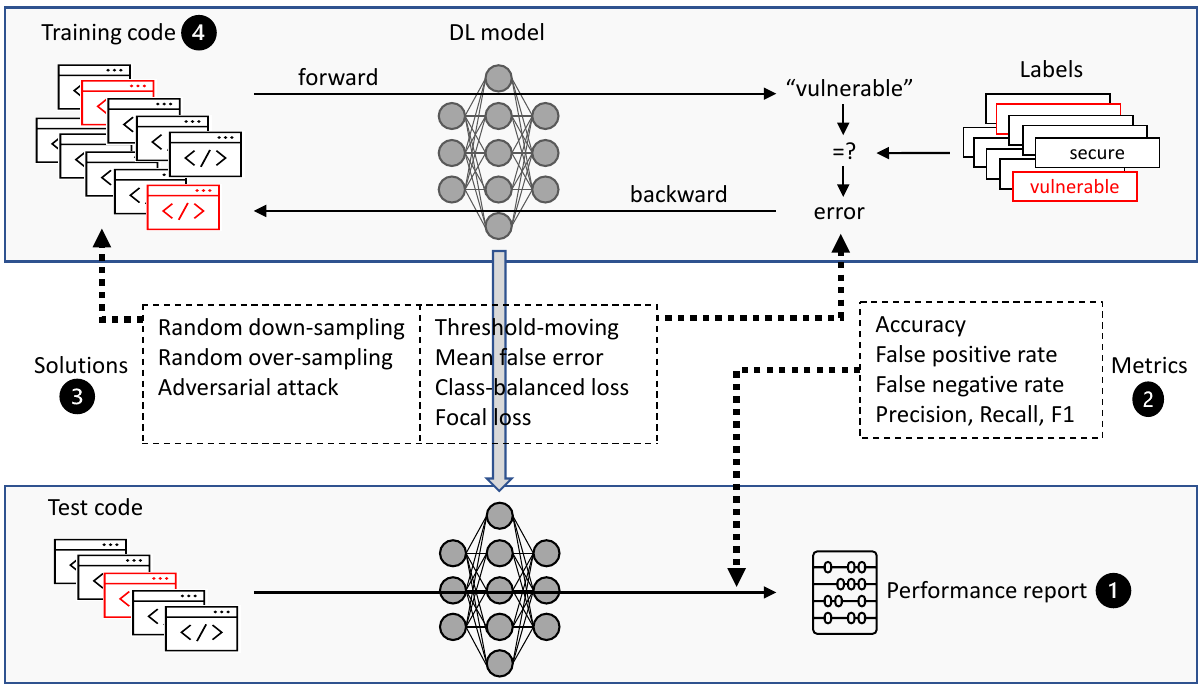}
    \caption{Overview of the empirical study. During the training procedure (top), a foundation DL model is fine-tuned to minimize the error between predicted and ground truth labels. In the test time (bottom), the trained model is used to make predictions on test data. \ding{182}\ding{183}\ding{184}\ding{185} refer to four research questions.}
    \label{fig:overview}
\end{figure}

Figure~\ref{fig:overview} gives an overview of our study design. In total, four research questions are framed: 

\noindent\textbf{RQ1:} How does imbalance impact model performance in vulnerability detection?

\noindent\textbf{RQ2:} Which metrics are appropriate for evaluating detection models?

\noindent\textbf{RQ3:} How effective are existing solutions in mitigating imbalance in vulnerability detection?

\noindent\textbf{RQ4:} What external factors may hinder the solutions to work?

\subsection{Sketch of DL-based Vulnerability Detection Task}
\label{subsec:problem}
Generally, deep learning approaches formalize the vulnerability detection task as a binary classification problem, i.e., identifying a given code sample as secure (the label is 0) or vulnerable (the label is 1)~\cite{devign2019zhou}. Formally, let $\mathcal{D}=\left\{x,y\right\}$ be a source code set with samples, where $x\in X$ and $y\in Y$ represent a source code sample and its corresponding label ($Y=\left\{0, 1\right\}$), respectively. $f_\theta$ is such a binary classifier parameterized by $\theta$ that maps the data space $X$ to the label space $Y$. The training procedure of the model is to find the optimal $\theta^*$ that minimizes the error between prediction and ground truth as shown in Figure~\ref{fig:overview}. Formally, the optimization objective is defined by:
\begin{equation}
\label{equ:erm}
    \theta^*= \argmin\limits_{\theta\in\Theta}\ell\left(f_\theta,X,Y\right)
\end{equation}
\noindent where $\ell$ is a loss function that captures the error between the predicted labels by $f_\theta$ and ground truth labels. The loss function can be in different forms, such as cross-entropy loss, mean squared error, and hinge embedding loss. For vulnerability detection, the cross-entropy loss is mostly applied and in this form, the objective becomes to minimize the cross entropy over all samples: 
\begin{equation}
\label{equ:loss}
    \ell=\frac{1}{N}\sum\limits_{i=1}^{N}CE\left(p_i,y_i\right)
\end{equation}
\noindent where 
\begin{equation}
\label{equ:ce}
    CE\left(p_i,y_i\right)=\begin{cases}-\log\left(p_i\right)&y_i=1\\-\log\left(1-p_i\right)&y_i=0\end{cases}
\end{equation}

\noindent Note that in Eq.~(\ref{equ:loss}), the loss is an average over all samples. That is, all samples are equally treated regardless of being secure or vulnerable, which is reasonable when data are evenly distributed in two types. However, in reality, the number of vulnerable code is usually much less than secure ones, which causes the imbalance issue for training. 

Remarkably, let $p$ ($0\leq p\leq 1$) be the predicted probability of $x$ being vulnerable. In binary classifiers, usually, a decision threshold is set at 0.5. Namely, if $p>0.5$, $x$ is determined as vulnerable, otherwise secure. 

\subsection{Imbalance Solutions}
\label{subsec:solutions}
In total, we examine seven methods that are widely adopted in the literature to handle the imbalance problem. These methods cover both the data and model-level solutions. Note that data level (marked with *) solutions only carry out on the training set during the training time. Let $N_s$ and $N_v$ denote the number of secure and vulnerable code, respectively. $N_d=N_s-N_v$.

\textbf{*Random down-sampling}~\cite{Chris2003down}. $N_d$ secure code are randomly selected and removed from the training set. 

\textbf{*Random over-sampling}~\cite{yin2021learning}. As opposed to down-sampling, $N_d$ vulnerable code are randomly selected from and replicated to the training set.

\textbf{*Adversarial attack-based augmentation}~\cite{alert2022,zhang2020generating}. An advanced over-sampling method that generates new vulnerable code via adversarial attack. In this paper, we perform the variable renaming-based adversarial attack that changes the name of local variables. The advantage is that the generated code can pass the compiler and remain executable. To ensure that the substitute name is natural to software developers, we use the masked language prediction function of CodeBERT~\cite{codebert2020} to produce the candidates. 

\textbf{Threshold-moving}~\cite{He2013thre}, also known as threshold-tuning, post-scaling, and thresholding. It adjusts the decision threshold and is applied during the test time. Concretely, the model is trained using the training set and then used to predict the probabilities of samples in the validation set. Given a candidate threshold set, the performance on the validation set is evaluated on each candidate and the threshold producing the best performance is selected as the optimal one for the prediction on the test set. In this paper, we set the candidate threshold to range from 0 to 1 with a 0.01 interval.

\textbf{Mean false error (MFE) loss}~\cite{seperated2016wang}. The concept comes from the ``false positive rate'' and ``false negative rate''. The goal is to make the loss more sensitive to the errors caused by the minority set by computing the loss on different sets separately. The original base loss is the mean squared error for image classification and document classification tasks. We adapt it to the cross entropy error to fit vulnerability detection models. Formally, 
\begin{equation}
\label{equ:mfe}
   \ell_{mfe}=\frac{1}{N_s}\sum\limits_{i=1}^{N_s}CE\left(p_i,y_i\right)+\frac{1}{N_v}\sum\limits_{j=1}^{N_v}CE\left(p_j,y_j\right)
\end{equation}

\textbf{Class-balanced (CB) loss}~\cite{cbl2019cui}. This method addresses the imbalance issue by introducing the effect number that refers to the expected volume of samples of a given set. Formally,
\begin{equation}
\label{equ:cbl}
\ell_{cb}=\frac{1}{N}\sum\limits_{i=1}^{N}\frac{1-\beta}{1-\beta^{N_{y_i}}}CE\left(p_i,y_i\right)
\end{equation}
\noindent where $N_{y_i}=N_v$ if $y_i=1$ otherwise $N_s$. As recommended~\cite{cbl2019cui}, we set $\beta=0.9999$.

\textbf{Focal loss (FL)}~\cite{focal2020lin}. The idea is to put more focus on hard, misclassified samples meanwhile reduce the loss for well-classified samples. For instance, given three vulnerable code $x_1$, $x_2$, and $x_3$, the model predicts them as vulnerable with 0.9, 0.5, and 0.2 probability, respectively. $x_1$ is well-classified. $x_2$ is hard to classify. $x_3$ is misclassified as secure. Formally,
\begin{equation}
\label{equ:fl}
\ell_{fl}=-\frac{1}{N}\sum\limits_{i=1}^{N}\left(1-p_t\right)^\gamma\log\left(p_t\right)
\end{equation}
\noindent where
\begin{equation}
\label{equ:cfl}
p_t=\begin{cases}p_i&y_i=1\\1-p_i&y_i=0\end{cases}
\end{equation}
\noindent when $\gamma=0$, Eq.~(\ref{equ:fl}) is equivalent to Eq.~(\ref{equ:loss}). As recommended~\cite{focal2020lin}, we set $\gamma=2$.

\subsection{Evaluation Metrics}
\label{subsec:metrics}
We investigate all the following six metrics which have been used in different papers to evaluate software vulnerability detection models~\cite{saikat2022yet,codebert2020,graphcodebert2020,devign2019zhou,vuldeepecker2018li}.

\textbf{Accuracy}~\cite{codebert2020,graphcodebert2020,devign2019zhou} is the percentage of samples that are correctly classified by a model. This metric may give a fake good performance. For instance, if a test set has 100 code where only one is vulnerable. A model classifies all samples as secure, so its accuracy is 99\%, which is nearly perfect. However, this model is useless as a detection model.

\textbf{False positive rate (FPR)}~\cite{vuldeepecker2018li} measures the ratio of misclassified secure code to the total number of secure samples. A low value means the model learns very well from secure code.

\textbf{False negative rate (FNR)}~\cite{vuldeepecker2018li} computes the ratio of misclassified vulnerable code to the total number of vulnerable samples. This metric focuses on the ability to figure out vulnerable code. A low value indicates a strong ability.

\textbf{Precision}~\cite{devign2019zhou}, also known as positive predictive rate, is the fraction of correctly classified vulnerable code among samples classified as vulnerable. 

\textbf{Recall}~\cite{devign2019zhou}, the opposite of FNR, is the fraction of correctly classified vulnerable code among all vulnerable samples. In practice, precision and recall are often in tension. Improving precision will cause recall to decay, and vice versa. 

\textbf{F1-measure (F1)}~\cite{devign2019zhou} is defined as the harmonic mean of precision and recall. It balances the importance between precision and recall.

\section{Experimental Setup}
All experiments were conducted on a high-performance computer (HPC) cluster and each cluster node runs a 2.20GHZ Intel Xeon Silver 4210 GPU with an NVIDIA Tesla V100 32G GPU. Models are trained and tested using the PyTorch 1.7.1 framework with CUDA 10.1. 

\subsection{Datasets}
\label{subsec:data}
As listed in Table~\ref{tab:datasets}, nine function-level datasets from three open-source repositories on GitHub are considered in the experiments. All the datasets are C-language and the related projects are popular among software developers. Devign~\cite{devignData} is provided by Zhou~\emph{et al.}~\cite{zhou2019devign} and consists of two datasets collected from the FFmpeg~\cite{ffmpegweb} and QEMU~\cite{qemuweb} projects, respectively. Labels of source code are manually annotated by professional security researchers. Lin2018~\cite{lin2018data}\footnote{Notice: the number of data is a bit different from the original paper in~\cite{lin2018cross} because we remove empty source code files from the provided datasets. Empty files cause compiling bugs and degrade the model performance.} includes six datasets from Asterisk~\cite{asteriskweb}, FFmpeg~\cite{ffmpegweb}, LibPNG~\cite{libpngweb}, LibTIFF~\cite{libtiffweb}, Pidgin~\cite{Pidginweb}, and VLC media player~\cite{vlcweb}, respectively. For each project, source code are manually labeled by Lin~\emph{et al.} according to the CVE and NVD records. CodeXGLUE~\cite{codexgluerepo} provides a mixture version of two datasets from FFmpeg and QEMU in Devign.

\begin{table}[ht]
\caption{Datasets overview. IR: imbalance ratio ($\frac{\#Secure}{\#Vulnerable}$).}
\centering
\label{tab:datasets}
\resizebox{\textwidth}{!}{%
\begin{tabular}{|l|l|l|r|r|r|r|}
\hline
\textbf{Source} & \textbf{Project} & \textbf{Project description} & \textbf{\#Vulnerable} & \textbf{\#Secure} & \textbf{\#Total} & \textbf{IR} \\ \hline
 & FFmpeg & A cross-platform to record, convert and stream audio and video. & 4,981 & 4,788 & 9,769 & 0.96 \\ \cline{2-7} 
\multirow{-2}{*}{Devign} & QEMU & A generic and open source machine emulator and virtualizer. & 7,479 & 10,070 & 17,549 & 1.35 \\ \hline
 & Asterisk & A framework for building communications applications. & 56 & 17,070 & 17,126 & 304.82 \\ \cline{2-7} 
 & FFmpeg & A cross-platform to record, convert and stream audio and video. & 213 & 5,550 & 5,763 & 26.06 \\ \cline{2-7} 
 & LibPNG & Official PNG reference. & 44 & 577 & 621 & 13.11 \\ \cline{2-7} 
 & LibTIFF & TIFF library and utilities. & 96 & 731 & 827 & 7.61 \\ \cline{2-7} 
 & {\color[HTML]{333333} Pidgin} & A multi-platform instant messaging client. & 29 & 8,612 & 8,641 & 296.97 \\ \cline{2-7} 
\multirow{-6}{*}{Lin2018} & VLC & A cross-platform multimedia player. & 43 & 6,113 & 6,156 & 142.16 \\ \hline
CodeXGLUE & Devign & Mixture of FFmpeg and QEMU. & 12,460 & 14,858 & 27,318 & 1.19 \\ \hline
\end{tabular}%
}
\end{table}

To have a closer view of the vulnerabilities existing in these datasets, we provide Table~\ref{tab:vulist}. In total, 25 vulnerabilities are included and a certain vulnerability may have more than one CVE record with different CVSS scores, e.g., Bypass a restriction or similar. 

\begin{table}[ht]
\caption{List of vulnerabilities. The common vulnerability scoring system (CVSS) score measures the severity of a certain vulnerability type.}
\label{tab:vulist}
\resizebox{.97\textwidth}{!}{%
\begin{tabular}{lll|lll}
\hline
\textbf{ID} & \textbf{Vulnerability Type} & \textbf{CVSS} & \textbf{ID} & \textbf{Vulnerability Type} & \textbf{CVSS} \\ \hline
1 & Bypass a restriction or similar & 4.3 - 7.5 & 14 & Execute Code & 5.8 - 9.3 \\
2 & Cross Site Scripting & 4.3 & 15 & Execute Code Gain privileges & 6.5 \\
3 & Denial Of Service & 2.6 - 7.8 & 16 & Execute Code Memory corruption & 6.8 \\
4 & Denial Of Service Execute Code & 6.8 - 9.3 & 17 & Execute Code Memory corruption Obtain Information & 6.8 \\
5 & Denial Of Service Execute Code Memory corruption & 6.8 - 10.0 & 18 & Execute Code Overflow & 6.0 - 10.0 \\
6 & Denial Of Service Execute Code Overflow & 6.5 - 10.0 & 19 & Execute Code Overflow Bypass a restriction or similar & 6.8 \\
7 & Denial Of Service Execute Code Overflow Memory corruption & 6.8 & 20 & Execute Code Overflow Memory corruption & 6.8 \\
8 & Denial Of Service Memory corruption & 6.8 & 21 & Gain privileges & 9.0 \\
9 & Denial Of Service Obtain Information & 4.3 - 10 & 22 & Obtain Information & 4.3 - 5.0 \\
10 & Denial Of Service Overflow & 2.6 - 9.3 & 23 & Overflow & 4.3 - 10 \\
11 & Denial Of Service Overflow Memory corruption & 4.3 - 6.8 & 24 & Overflow Memory corruption & 5.0 \\
12 & Denial Of Service Overflow Obtain Information & 5.8 & 25 & Unspecified & 4.3 - 10.0 \\
13 & Directory traversal & 5.8 - 9.3 &  &  &  \\ \hline
\end{tabular}%
}
\end{table}

\subsection{Models}
\label{subsec:models}

Two SOTA foundation models, CodeBERT~\cite{codebert2020} and GraphCodeBERT~\cite{graphcodebert2020}, for natural language and programming language, are leveraged in this paper. Both models follow BERT~\cite{bert2019devlin} and use multi-layer bidirectional Transformer~\cite{NIPS2017_3f5ee243} as the backbone. CodeBERT is pre-trained on 2.1M bimodal data and 6.4M unimodal codes. GraphCodeBERT is pre-trained on the CodeSearchNet~\cite{husain2019codesearchnet} dataset consisting of 2.3M functions paired with natural language descriptions. The main difference between CodeBERT and GraphCodeBERT is that the source code in CodeBERT is represented as a sequence of tokens, while GraphCodeBERT takes the data flow of source code as its input. 

Our implementation is adapted from the GitHub repositories provided by CodeXGLUE\footnote{\url{https://github.com/microsoft/CodeXGLUE}} for CodeBERT and by Microsoft\footnote{\url{https://github.com/microsoft/CodeBERT}} for GraphCodeBERT, respectively. The base models for fine-tuning are loaded from Hugging Face\footnote{\url{https://huggingface.co/microsoft/codebert-base}}\footnote{\url{https://huggingface.co/microsoft/graphcodebert-base}} from Hugging Face.

\subsection{Training}
\label{subsec:training}
Each model is fine-tuned 50 epochs and the ``best'' one is saved for evaluation. For reproduction, we follow the default setting in original implementations to set the random seed at 123456. In each dataset, we proportionally (8:1:1) split the dataset into a training set, a validation set, and a test set (the training and validation sets are involved in the training procedure, and the test set is only for testing.). Vulnerable and secure code are randomly divided into these three sets with the same imbalance ratio as in Table~\ref{tab:datasets}. 

\section{Results}
\label{sec:results}

\subsection{RQ1: Influence of Imbalance in Vulnerability Detection} 
\label{subsec:rq1}
\emph{Experiments.} We train CodeBERT and GraphCodeBERT with default settings, ignoring the imbalance. For each trained model, we check if the imbalance causes a bias towards secure code by comparing loss and accuracy across individual sets. 

\emph{Results.} Table~\ref{tab:rq1} shows the results. Regardless of the model and evaluation metric, the model performs better on the secure set than on the vulnerable one. Considering the accuracy, except for the FFmpeg dataset in Devign, both models achieve higher accuracy on secure code. Particularly, 100\% secure code can be perfectly identified in several datasets, such as LibPNG, LibTIFF, Pidgin, and VLC. However, the performance in identifying vulnerable code is less satisfying. For instance, in Asterisk from Lin2018, GraphCodeBERT can correctly identify all secure code but only 44.44\% vulnerable code. On the other hand, with respect to the loss between prediction and ground truth, the loss on secure code is, in general, much lower than on vulnerable code. For instance, on average, CodeBERT has no loss on each secure code for Pidgin from Lin2018, but 0.44 on the vulnerable one. This indicates that during the training procedure, the model tends to learn more from the secure code. When summing over all code, the imbalance makes it worse, e.g., GraphCodeBERT has, in total, 0.39 loss on secure code but 13.18 on vulnerable one. The reason is that, during the training procedure, the loss is calculated as an average (Eq.~(\ref{equ:loss})) or sum over all samples (both the majority secure and minority vulnerable). This methodology weakens the influence of vulnerable code and gives ``fake'' feedback to the training that the model is performing well, which is the essence of the imbalance issue.

\begin{table}[t]
\caption{Model accuracy and loss on secure and vulnerable code. \textbf{Baseline}: model accuracy on all code. The best performance is highlighted.}
\centering
\label{tab:rq1}
\resizebox{.95\textwidth}{!}{%
\begin{tabular}{llccccccc}
\hline
 & \multicolumn{1}{c}{} & \multicolumn{3}{c}{\textbf{Accuracy (\%)}} & \multicolumn{2}{c}{\textbf{Total Loss}} & \multicolumn{2}{c}{\textbf{Average loss}} \\ \cline{3-9} 
\multirow{-2}{*}{\textbf{Source}} & \multicolumn{1}{c}{\multirow{-2}{*}{\textbf{Project}}} & \textbf{Baseline} & \textbf{Vulnerable} & \textbf{Secure} & \textbf{Vulnerable} & \textbf{Secure} & \textbf{Vulnerable} & \textbf{Secure} \\ \hline
\multicolumn{9}{c}{\textbf{CodeBERT}} \\ \hline
 & FFmpeg & 56.71 & \cellcolor[HTML]{C0C0C0}62.17 & 51.04 & 537.37 & \cellcolor[HTML]{C0C0C0}449.77 & 0.72 & \cellcolor[HTML]{C0C0C0}0.63 \\
\multirow{-2}{*}{Devign} & QEMU & 64.31 & 40.96 & \cellcolor[HTML]{C0C0C0}81.67 & 1361.10 & \cellcolor[HTML]{C0C0C0}583.43 & 1.21 & \cellcolor[HTML]{C0C0C0}0.39 \\ \hline
 & Asterisk & 99.77 & 44.44 & \cellcolor[HTML]{C0C0C0}99.96 & 33.75 & \cellcolor[HTML]{C0C0C0}3.18 & 3.75 & \cellcolor[HTML]{C0C0C0}0.00 \\
 & FFmpeg & 97.46 & 36.36 & \cellcolor[HTML]{C0C0C0}99.88 & 124.56 & \cellcolor[HTML]{C0C0C0}2.92 & 3.77 & \cellcolor[HTML]{C0C0C0}0.00 \\
 & LibPNG & 95.83 & 50.00 & \cellcolor[HTML]{C0C0C0}100.00 & 14.25 & \cellcolor[HTML]{C0C0C0}0.65 & 1.78 & \cellcolor[HTML]{C0C0C0}0.01 \\
 & LibTIFF & 88.89 & 53.33 & \cellcolor[HTML]{C0C0C0}93.69 & 49.34 & \cellcolor[HTML]{C0C0C0}30.54 & 3.29 & \cellcolor[HTML]{C0C0C0}0.28 \\
 & Pidgin & 99.85 & 60.00 & \cellcolor[HTML]{C0C0C0}100.00 & 17.14 & \cellcolor[HTML]{C0C0C0}0.08 & 3.43 & \cellcolor[HTML]{C0C0C0}0.00 \\
\multirow{-6}{*}{Lin2018} & VLC & 99.89 & 85.71 & \cellcolor[HTML]{C0C0C0}100.00 & 3.07 & \cellcolor[HTML]{C0C0C0}0.08 & 0.44 & \cellcolor[HTML]{C0C0C0}0.00 \\ \hline
CodeXGLUE & Devign & 61.49 & 31.95 & \cellcolor[HTML]{C0C0C0}86.59 & 1716.89 & \cellcolor[HTML]{C0C0C0}531.24 & 1.37 & \cellcolor[HTML]{C0C0C0}0.36 \\ \hline
\multicolumn{9}{c}{\textbf{GraphCodeBERT}} \\ \hline
 & FFmpeg & 56.99 & \cellcolor[HTML]{C0C0C0}74.87 & 38.39 & \cellcolor[HTML]{C0C0C0}437.05 & 559.02 & \cellcolor[HTML]{C0C0C0}0.58 & 0.78 \\
\multirow{-2}{*}{Devign} & QEMU & 65.38 & 52.98 & \cellcolor[HTML]{C0C0C0}74.59 & 1001.18 & \cellcolor[HTML]{C0C0C0}776.03 & 0.89 & \cellcolor[HTML]{C0C0C0}0.51 \\ \hline
 & Asterisk & 99.81 & 44.44 & \cellcolor[HTML]{C0C0C0}100.00 & 31.18 & \cellcolor[HTML]{C0C0C0}0.90 & 3.46 & \cellcolor[HTML]{C0C0C0}0.00 \\
 & FFmpeg & 97.35 & 51.52 & \cellcolor[HTML]{C0C0C0}99.16 & 134.98 & \cellcolor[HTML]{C0C0C0}37.00 & 4.09 & \cellcolor[HTML]{C0C0C0}0.04 \\
 & LibPNG & 96.88 & 62.50 & \cellcolor[HTML]{C0C0C0}100.00 & 16.16 & \cellcolor[HTML]{C0C0C0}0.68 & 2.02 & \cellcolor[HTML]{C0C0C0}0.01 \\
 & LibTIFF & 92.06 & 33.33 & \cellcolor[HTML]{C0C0C0}100.00 & 47.81 & \cellcolor[HTML]{C0C0C0}1.31 & 3.19 & \cellcolor[HTML]{C0C0C0}0.01 \\
 & Pidgin & 99.85 & 60.00 & \cellcolor[HTML]{C0C0C0}100.00 & 13.18 & \cellcolor[HTML]{C0C0C0}0.39 & 2.64 & \cellcolor[HTML]{C0C0C0}0.00 \\
\multirow{-6}{*}{Lin2018} & VLC & 99.68 & 57.14 & \cellcolor[HTML]{C0C0C0}100.00 & 11.77 & \cellcolor[HTML]{C0C0C0}1.26 & 1.68 & \cellcolor[HTML]{C0C0C0}0.00 \\ \hline
CodeXGLUE & Devign & 62.81 & 58.25 & \cellcolor[HTML]{C0C0C0}66.69 & 1047.34 & \cellcolor[HTML]{C0C0C0}806.85 & 0.83 & \cellcolor[HTML]{C0C0C0}0.55 \\ \hline
\end{tabular}%
}
\end{table}

\emph{Answer}: The imbalance encourages a DL model to gain more knowledge from secure code, which leads to poor performance on detecting vulnerable code, e.g., 44.44\% accuracy (55.56\% false negative rate).

\subsection{RQ2: Analysis of Evaluation Metrics}
\label{subsec:rq2}

\emph{Experiments.} In the default setting of CodeBERT and GraphCodeBERT, for each trained model, different metrics are used to evaluate the model performance.

\emph{Results.} Table~\ref{tab:rq2} lists the model performance. From the perspective of identifying vulnerable code, accuracy and FPR are not informative enough and can be misleading. For instance, Table~\ref{tab:rq1} shows that CodeBERT only successfully detects 44.44\% vulnerable code in Asterisk from Lin2018. However, the output accuracy is almost perfect at 99.77\%. The overall accuracy hides the actual performance of detecting vulnerable code. FPR only considers the detection of secure code, which misses the main purpose of vulnerable detection, namely to identify vulnerabilities at an early stage. Recall (the opposite of FNR) is equivalent to the individual accuracy on the vulnerable code in Table~\ref{tab:rq1} and can tell how the model identifies vulnerable code. However, recall ignores the secure code. Precision covers this shortage by including misclassified secure code. If one only cares about detecting vulnerable code and ignores the cost of manually filtering secure code afterward, recall is the best option. If one wishes to have fewer errors in the identified vulnerable code, precision can be taken. As a balanced version between precision and recall, F1 can be used when an overall score is preferred. 

\begin{table}[ht]
\caption{Model performance (\%) using different evaluation metrics. Accuracy is the default metric of CodeBERT and GraphCodeBERT.}
\centering
\label{tab:rq2}
\resizebox{.8\textwidth}{!}{%
\begin{tabular}{llcccccc}
\hline
\textbf{Source} & \textbf{Project} & \textbf{Accuracy} & \textbf{FPR} & \textbf{FNR} & \textbf{Precision} & \textbf{Recall} & \textbf{F1} \\ \hline
\multicolumn{8}{c}{\textbf{CodeBERT}} \\ \hline
\multirow{2}{*}{Devign} & FFmpeg & 56.71 & 48.96 & 37.83 & 56.92 & 62.17 & 59.42 \\
 & QEMU & 64.31 & 18.33 & 59.04 & 62.42 & 40.96 & 49.46 \\ \hline
\multirow{6}{*}{Lin2018} & Asterisk & 99.77 & 0.04 & 55.56 & 80.00 & 44.44 & 57.14 \\
 & FFmpeg & 97.46 & 0.12 & 63.64 & 92.31 & 36.36 & 52.17 \\
 & LibPNG & 95.83 & 0.00 & 50.00 & 100.00 & 50.00 & 66.67 \\
 & LibTIFF & 88.89 & 6.31 & 46.67 & 53.33 & 53.33 & 53.33 \\
 & Pidgin & 99.85 & 0.00 & 40.00 & 100.00 & 60.00 & 75.00 \\
 & VLC & 99.89 & 0.00 & 14.29 & 100.00 & 85.71 & 92.31 \\ \hline
CodeXGLUE & Devign & 61.49 & 13.41 & 68.05 & 66.94 & 31.95 & 43.26 \\ \hline
\multicolumn{8}{c}{\textbf{GraphCodeBERT}} \\ \hline
\multirow{2}{*}{Devign} & FFmpeg & 56.99 & 61.61 & 25.13 & 55.83 & 74.87 & 63.96 \\
 & QEMU & 65.38 & 25.41 & 47.02 & 60.78 & 52.98 & 56.61 \\ \hline
\multirow{6}{*}{Lin2018} & Asterisk & 99.81 & 0.00 & 55.56 & 100.00 & 44.44 & 61.54 \\
 & FFmpeg & 97.35 & 0.84 & 48.48 & 70.83 & 51.52 & 59.65 \\
 & LibPNG & 96.88 & 0.00 & 37.50 & 100.00 & 62.50 & 76.92 \\
 & LibTIFF & 92.06 & 0.00 & 66.67 & 100.00 & 33.33 & 50.00 \\
 & Pidgin & 99.85 & 0.00 & 40.00 & 100.00 & 60.00 & 75.00 \\
 & VLC & 99.68 & 0.00 & 42.86 & 100.00 & 57.14 & 72.73 \\ \hline
CodeXGLUE & Devign & 62.81 & 33.31 & 41.75 & 59.77 & 58.25 & 59.00 \\ \hline
\end{tabular}%
}
\end{table}

\emph{Answer}: Precision, recall and F1 provide more informative and comprehensive insights on model performance than accuracy. FPR might be useful in some situations to limit the impact of false positives (e.g., static analysis), but precision can serve a similar purpose as a higher precision generally implies a lower FPR.    
 
\subsection{RQ3: Effectiveness of Solutions for Addressing Imbalance}
\label{subsec:rq3}
\emph{Experiments.} We train CodeBERT and GraphCodeBERT following the methodology of different solutions for handling the imbalance issue. Based on the selected evaluation metrics, precision, recall, and F1, by RQ2, the effectiveness of solutions is investigated.

\emph{Results.} Table~\ref{tab:rq3-code} and Table~\ref{tab:rq3-graph} show the results on CodeBERT and GraphCodeBERT, respectively. Note that in FFmpeg, Devign, the number of vulnerable programs (4981) is greater than the secure one (4788), thus, no re-sampling-based solutions are applied. Regardless of the dataset, model, and evaluation metric, random down-sampling performs the worst since massive information about the secure code is eliminated. In particular, when the imbalance ratio is high and the data size is small (e.g., Asterisk from Lin2018), the remaining data is insufficient to support the model training. The focal loss stands out as the optimal choice for improving precision. The reason is that focal loss puts more effort into hard and misclassified samples during the training procedure whether those samples are vulnerable or secure. Thus, the model can more precisely predict a code sample to be vulnerable or secure. Two model-level solutions, MFE and CB, are the worst regarding precision. The reason is that, the methodology of these two solutions is to put relatively more attention to the vulnerable code during the training procedure, thus, more vulnerable code should be correctly identified than the baseline. This is confirmed by the results of recall where both solutions outperform the others. While in this case, the focal loss gains low recall. With respect to the overall performance F1, random over-sampling seems to be the best in most cases for both models. 

\begin{table}[ht]
\caption{CodeBERT trained using different solutions for imbalance issues. \textbf{Baseline}: the default setting. \textbf{Down-R}: random down-sampling. \textbf{Over-R}: random over-sampling. \textbf{Over-A}: adversarial attack-based augmentation. For each dataset, the best solution under a given metric is highlighted.}
\label{tab:rq3-code}
\resizebox{\textwidth}{!}{%
\begin{tabular}{llcccccccc}
\hline
\textbf{Source} & \textbf{Project} & \textbf{Baseline} & \textbf{Down-R} & \textbf{Over-R} & \textbf{Over-A} & \textbf{Thresholding} & \textbf{MFE} & \textbf{CB} & \textbf{FL} \\ \hline
\multicolumn{10}{c}{\textbf{Precision}} \\ \hline
Devign & FFmpeg & 56.92 & - & - & - & 0.00 & 67.72 & 63.14 & \cellcolor[HTML]{C0C0C0}88.24 \\
 & QEMU & 62.42 & 60.51 & 62.33 & 71.12 & 0.00 & 55.36 & 55.17 & \cellcolor[HTML]{C0C0C0}93.04 \\ \hline
 & Asterisk & 80.00 & 0.00 & 75.00 & \cellcolor[HTML]{C0C0C0}100.00 & 0.00 & 66.67 & 54.55 & \cellcolor[HTML]{C0C0C0}100.00 \\
 & FFmpeg & 92.31 & 3.23 & 76.19 & 65.38 & \cellcolor[HTML]{C0C0C0}100.00 & 34.29 & 38.71 & 90.00 \\
 & LibPNG & 100.00 & 66.67 & \cellcolor[HTML]{C0C0C0}100.00 & 85.71 & 0.00 & 62.50 & 70.00 & \cellcolor[HTML]{C0C0C0}100.00 \\
 & LibTIFF & 53.33 & 66.67 & 60.00 & \cellcolor[HTML]{C0C0C0}80.00 & 72.73 & 57.14 & 53.33 & \cellcolor[HTML]{C0C0C0}80.00 \\
 & Pidgin & 100.00 & 0.00 & \cellcolor[HTML]{C0C0C0}100.00 & 75.00 & \cellcolor[HTML]{C0C0C0}100.00 & 60.00 & 0.00 & 0.00 \\
\multirow{-6}{*}{Lin2018} & VLC & 100.00 & 0.00 & \cellcolor[HTML]{C0C0C0}100.00 & \cellcolor[HTML]{C0C0C0}100.00 & \cellcolor[HTML]{C0C0C0}100.00 & 0.00 & 0.00 & 0.00 \\ \hline
CodeXGLUE & Devign & 66.94 & 59.52 & 62.03 & 62.72 & \cellcolor[HTML]{C0C0C0}100.00 & 58.38 & 59.82 & 85.49 \\ \hline
\multicolumn{10}{c}{\textbf{Recall}} \\ \hline
 & FFmpeg & 62.17 & - & - & - & 0.00 & 31.42 & \cellcolor[HTML]{C0C0C0}47.86 & 4.01 \\
\multirow{-2}{*}{Devign} & QEMU & 40.96 & 50.49 & 49.07 & 29.39 & 0.00 & 71.33 & \cellcolor[HTML]{C0C0C0}75.96 & 9.53 \\ \hline
 & Asterisk & 44.44 & 0.00 & 66.67 & 44.44 & 0.00 & 44.44 & \cellcolor[HTML]{C0C0C0}66.67 & 33.33 \\
 & FFmpeg & 36.36 & 3.03 & 48.48 & 51.52 & 36.36 & \cellcolor[HTML]{C0C0C0}72.73 & \cellcolor[HTML]{C0C0C0}72.73 & 54.55 \\
 & LibPNG & 50.00 & 25.00 & 62.50 & 75.00 & 0.00 & 62.50 & \cellcolor[HTML]{C0C0C0}87.50 & 62.50 \\
 & LibTIFF & 53.33 & 40.00 & 40.00 & 26.67 & \cellcolor[HTML]{C0C0C0}53.33 & \cellcolor[HTML]{C0C0C0}53.33 & \cellcolor[HTML]{C0C0C0}53.33 & 26.67 \\
 & Pidgin & 60.00 & 0.00 & \cellcolor[HTML]{C0C0C0}60.00 & \cellcolor[HTML]{C0C0C0}60.00 & \cellcolor[HTML]{C0C0C0}60.00 & \cellcolor[HTML]{C0C0C0}60.00 & 0.00 & 0.00 \\
\multirow{-6}{*}{Lin2018} & VLC & 85.71 & 0.00 & \cellcolor[HTML]{C0C0C0}85.71 & 57.14 & \cellcolor[HTML]{C0C0C0}85.71 & 0.00 & 0.00 & 0.00 \\ \hline
CodeXGLUE & Devign & 31.95 & 47.33 & 52.59 & 45.98 & 5.02 & \cellcolor[HTML]{C0C0C0}53.55 & 47.81 & 13.15 \\ \hline
\multicolumn{10}{c}{\textbf{F1}} \\ \hline
 & FFmpeg & 59.42 & - & - & - & 0.00 & 42.92 & \cellcolor[HTML]{C0C0C0}54.45 & 7.67 \\
\multirow{-2}{*}{Devign} & QEMU & 49.46 & 55.05 & 54.91 & 41.59 & 0.00 & 62.33 & \cellcolor[HTML]{C0C0C0}63.92 & 17.29 \\ \hline
 & Asterisk & 57.14 & 0.00 & \cellcolor[HTML]{C0C0C0}70.59 & 61.54 & 0.00 & 53.33 & 60.00 & 50.00 \\
 & FFmpeg & 52.17 & 3.13 & 59.26 & 57.63 & 53.33 & 46.60 & 50.53 & \cellcolor[HTML]{C0C0C0}67.92 \\
 & LibPNG & 66.67 & 36.36 & 76.92 & \cellcolor[HTML]{C0C0C0}80.00 & 0.00 & 62.50 & 77.78 & 76.92 \\
 & LibTIFF & 53.33 & 50.00 & 48.00 & 40.00 & \cellcolor[HTML]{C0C0C0}61.54 & 55.17 & 53.33 & 40.00 \\
 & Pidgin & 75.00 & 0.00 & \cellcolor[HTML]{C0C0C0}75.00 & 66.67 & \cellcolor[HTML]{C0C0C0}75.00 & 60.00 & 0.00 & 0.00 \\
\multirow{-6}{*}{Lin2018} & VLC & 92.31 & 0.00 & \cellcolor[HTML]{C0C0C0}92.31 & 72.73 & \cellcolor[HTML]{C0C0C0}92.31 & 0.00 & 0.00 & 0.00 \\ \hline
CodeXGLUE & Devign & 43.26 & 52.73 & \cellcolor[HTML]{C0C0C0}56.92 & 53.06 & 9.56 & 55.86 & 53.14 & 22.79 \\ \hline
\end{tabular}%
}
\end{table}

\begin{table}[ht]
\caption{GraphCodeBERT trained using different solutions for imbalance issue. \textbf{Baseline}: the default setting. \textbf{Down-R}: random down-sampling. \textbf{Over-R}: random over-sampling. \textbf{Over-A}: adversarial attack-based augmentation. For each dataset, the best solution under a given metric is highlighted.}
\label{tab:rq3-graph}
\resizebox{\textwidth}{!}{%
\begin{tabular}{llcccccccc}
\hline
\textbf{Source} & \textbf{Project} & \textbf{Baseline} & \textbf{Down-R} & \textbf{Over-R} & \textbf{Over-A} & \textbf{Thresholding} & \textbf{MFE} & \textbf{CB} & \textbf{FL} \\ \hline
\multicolumn{10}{c}{\textbf{Precision}} \\ \hline
Devign & FFmpeg & 55.83 & - & - & - & 0.00 & 59.48 & 56.76 & \cellcolor[HTML]{C0C0C0}86.36 \\
 & QEMU & 60.78 & 58.43 & 61.19 & 65.28 & \cellcolor[HTML]{C0C0C0}93.33 & 56.70 & 54.93 & 85.15 \\ \hline
 & Asterisk & 100.00 & 0.00 & 44.44 & \cellcolor[HTML]{C0C0C0}100.00 & \cellcolor[HTML]{C0C0C0}100.00 & 38.46 & 23.53 & \cellcolor[HTML]{C0C0C0}100.00 \\
 & FFmpeg & 70.83 & 24.53 & 70.97 & 70.37 & \cellcolor[HTML]{C0C0C0}75.00 & 28.00 & 17.39 & 62.50 \\
 & LibPNG & 100.00 & 66.67 & 83.33 & 85.71 & 0.00 & 66.67 & 66.67 & \cellcolor[HTML]{C0C0C0}100.00 \\
 & LibTIFF & 100.00 & 0.00 & \cellcolor[HTML]{C0C0C0}100.00 & 77.78 & 0.00 & 55.56 & 47.06 & \cellcolor[HTML]{C0C0C0}100.00 \\
 & Pidgin & 100.00 & 0.00 & \cellcolor[HTML]{C0C0C0}100.00 & 75.00 & 0.00 & 50.00 & 20.00 & 66.67 \\
\multirow{-6}{*}{Lin2018} & VLC & 100.00 & 0.00 & 83.33 & \cellcolor[HTML]{C0C0C0}100.00 & 0.00 & 75.00 & 75.00 & \cellcolor[HTML]{C0C0C0}100.00 \\ \hline
CodeXGLUE & Devign & 59.77 & 58.25 & 59.75 & 60.05 & \cellcolor[HTML]{C0C0C0}100.00 & 60.36 & 60.86 & 90.21 \\ \hline
\multicolumn{10}{c}{\textbf{Recall}} \\ \hline
 & FFmpeg & 74.87 & - & - & - & 0.00 & 52.41 & \cellcolor[HTML]{C0C0C0}65.11 & 5.08 \\
\multirow{-2}{*}{Devign} & QEMU & 52.98 & 54.94 & 53.07 & 37.67 & 6.23 & \cellcolor[HTML]{C0C0C0}64.02 & 62.51 & 15.32 \\ \hline
 & Asterisk & 44.44 & 0.00 & 44.44 & 44.44 & 33.33 & \cellcolor[HTML]{C0C0C0}55.56 & 44.44 & 22.22 \\
 & FFmpeg & 51.52 & 78.79 & 66.67 & 57.58 & 45.45 & \cellcolor[HTML]{C0C0C0}84.85 & \cellcolor[HTML]{C0C0C0}84.85 & 60.61 \\
 & LibPNG & 62.50 & \cellcolor[HTML]{C0C0C0}75.00 & 62.50 & \cellcolor[HTML]{C0C0C0}75.00 & 0.00 & \cellcolor[HTML]{C0C0C0}75.00 & \cellcolor[HTML]{C0C0C0}75.00 & 62.50 \\
 & LibTIFF & 33.33 & 0.00 & 40.00 & 46.67 & 0.00 & \cellcolor[HTML]{C0C0C0}66.67 & 53.33 & 26.67 \\
 & Pidgin & 60.00 & 0.00 & \cellcolor[HTML]{C0C0C0}80.00 & 60.00 & 0.00 & \cellcolor[HTML]{C0C0C0}80.00 & 60.00 & \cellcolor[HTML]{C0C0C0}80.00 \\
\multirow{-6}{*}{Lin2018} & VLC & 57.14 & 0.00 & 71.43 & \cellcolor[HTML]{C0C0C0}85.71 & 0.00 & \cellcolor[HTML]{C0C0C0}85.71 & \cellcolor[HTML]{C0C0C0}85.71 & 71.43 \\ \hline
CodeXGLUE & Devign & 58.25 & 55.70 & 56.89 & 54.50 & 4.54 & \cellcolor[HTML]{C0C0C0}61.04 & 57.85 & 13.94 \\ \hline
\multicolumn{10}{c}{\textbf{F1}} \\ \hline
 & FFmpeg & 63.96 & - & - & - & 0.00 & 55.72 & \cellcolor[HTML]{C0C0C0}60.65 & 9.60 \\
\multirow{-2}{*}{Devign} & QEMU & 56.61 & 56.63 & 56.84 & 47.77 & 11.69 & 60.14 & \cellcolor[HTML]{C0C0C0}58.48 & 25.96 \\ \hline
 & Asterisk & 61.54 & 0.00 & 44.44 & \cellcolor[HTML]{C0C0C0}61.54 & 50.00 & 45.45 & 30.77 & 36.36 \\
 & FFmpeg & 59.65 & 37.41 & \cellcolor[HTML]{C0C0C0}68.75 & 63.33 & 56.60 & 42.11 & 28.87 & 61.54 \\
 & LibPNG & 76.92 & 70.59 & 71.43 & \cellcolor[HTML]{C0C0C0}80.00 & 0.00 & 70.59 & 70.59 & 76.92 \\
 & LibTIFF & 50.00 & 0.00 & 57.14 & 58.33 & 0.00 & \cellcolor[HTML]{C0C0C0}60.61 & 50.00 & 42.11 \\
 & Pidgin & 75.00 & 0.00 & \cellcolor[HTML]{C0C0C0}88.89 & 66.67 & 0.00 & 61.54 & 30.00 & 72.73 \\
\multirow{-6}{*}{Lin2018} & VLC & 72.73 & 0.00 & 76.92 & 92.31 & 0.00 & 80.00 & 80.00 & \cellcolor[HTML]{C0C0C0}83.33 \\ \hline
CodeXGLUE & Devign & 59.00 & 56.95 & 58.29 & 57.14 & 8.69 & \cellcolor[HTML]{C0C0C0}60.70 & 59.31 & 24.15 \\ \hline
\end{tabular}%
}
\end{table}

\emph{Answer}: No single existing solution is the best to address the imbalance issue across all evaluation metrics. Specifically, to focal loss is the best option for improving precision. MFE and CB shall be used for optimizing recall. Random over-sampling is the best option when focusing on the overall F1 performance. Nevertheless, the pursuit of a new, task-specific solution to address the imbalance issue remains imperative.   

\subsection{RQ4: Investigation of External Factors}
\label{subsec:rq4}
\emph{Experiments.} Based on Table~\ref{tab:rq3-code} and Table~\ref{tab:rq3-graph}, we dig into the prediction results to explore possible external factors.

\begin{figure}[ht]
    \centering
    \subfigure[LibTIFF]{\label{fig:vul_libtiff}
    \includegraphics[scale=0.38]{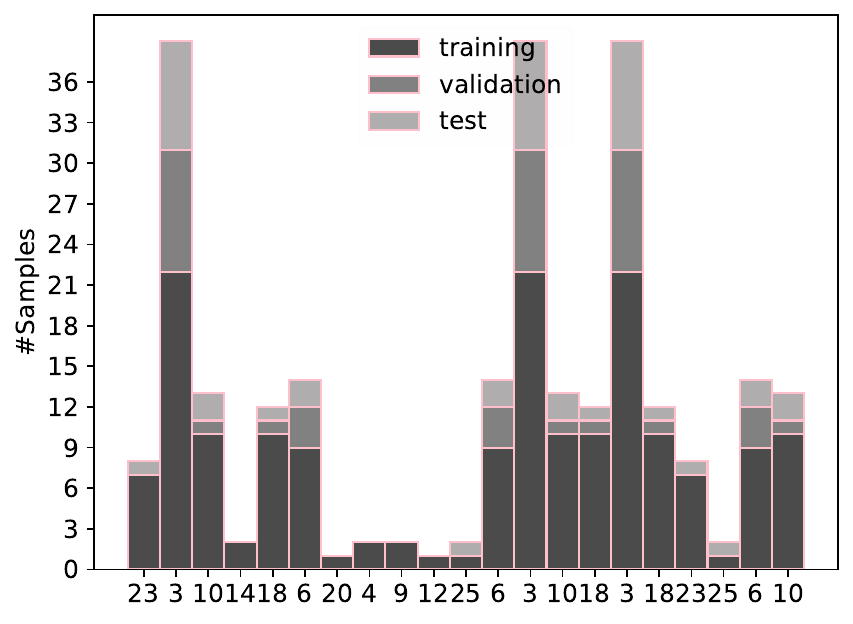}
    }
    \subfigure[Pidgin]{\label{fig:vul_pidgin}
    \includegraphics[scale=0.38]{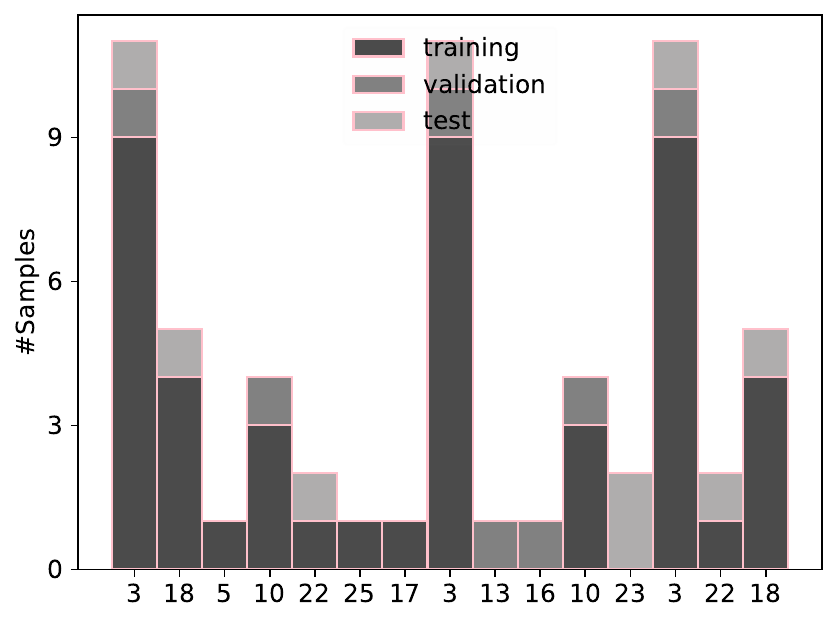}
    }
    \caption{Vulnerability type distribution in each split set (training, validation, and test). $x$-axis: vulnerability type ID (Please refer to Table~\ref{tab:vulist} for more details.). $y$-axis: number of samples in the corresponding set. Source: Lin2018.}
    \label{fig:vul}
\end{figure}

\emph{Results.} Note that in some cases, the model performance degrades after applying a solution. For instance, in Pidgin from Lin2018, CodeBERT with the default setting gains 100\% precision and 60\% recall, but 75\% precision and the same recall by adversarial attack-based augmentation. We found that by over-R, over-A, thresholding, and MFE, CodeBERT identifies the same vulnerable samples as the baseline, and the code with the Overflow vulnerability (ID 23 in Table~\ref{tab:vulist}) is always misclassified. This is because this vulnerability type does not appear in the training or validation sets (as shown in Figure~\ref{fig:vul_pidgin}) and the model cannot gain knowledge of this specific vulnerability type. Introducing more vulnerable samples can just cause the overfitting problem. Another case is in LibTIFF from Lin2018, the thresholding, MFE, and CB identify the same vulnerable samples as the baseline and miss five vulnerability types, Denial Of Service (ID 3,) Denial Of Service Execute Code Overflow (ID 6), Denial Of Service Overflow (ID 10), Execute Code Overflow (ID 18), and Overflow (ID 23). All these types are included in the training procedure (training and validation sets) (see Figure~\ref{fig:vul_libtiff}). All the trained models with or without solutions reach 53.33\% recall, these solutions can only increase the correctness of secure code because, by the corresponding training methodology, the model already reaches the limit of identifying certain types of vulnerability. In addition, thresholding tends to ruin the model entirely, such as FFmpeg from Devign and LibPNG from Lin2018, which is caused by the distribution shift between the validation set in the training time and the test set in the test time. Distribution shift~\cite{wilds2021} is a research topic per se and is not further explained in this paper.

\emph{Answer}: External factors including the absence of vulnerability types in the training time, inherent identification difficulty of certain vulnerability types, and the distribution shift in data should be considered when developing a new solution.

\subsection{Insights}
\label{subsec:insights}
\textbf{Selecting evaluation metrics:} In vulnerability detection, when selecting a metric to evaluate a model's performance, accuracy is the least suitable metric. Recall should be used if only the detection on vulnerable code matters. Precision should be selected if one wishes to have less secure code in identified vulnerable code. F1 can be considered from an overall perspective. 

\textbf{Designing solutions:} When designing a solution to address imbalance, one should consider the evaluation metric first. If the goal is to improve precision or recall, modifying the loss function is more efficient than manipulating training data. Minority over-sampling brings benefits to the overall evaluation. Vulnerability type and difficulty in data should be considered when a solution fails. 

\section{Conclusion}
\label{sec:conclusion}
This work studies the imbalance issue in software vulnerability detection. Seven solutions proposed in other domains are investigated on nine open-source datasets and two state-of-the-art deep learning models (CodeBERT and GraphCodeBERT). We found the defaulting setting of CodeBERT and GraphCodeBERT makes the training procedure focus more on the secure code, which causes a high false negative rate (e.g., 68.05\%). Existing solutions perform differently over various datasets and models, which calls for a new solution specifically for vulnerability detection. With the insights stated in the paper, this will be an interesting future work. Furthermore, we explore external factors like the vulnerability type distribution that should be aware of when designing such a new solution. There are many future research topics. The observations from this paper should be tested on other datasets and for other programming languages. Related to this, the observations should also be tested on other ML models other than CodeBERT and GraphCodeBERT. External factors, which can affect the performances, should be explored in more depth. This is particularly important if a solution is about to be deployed in practice.  

\clearpage
\bibliographystyle{splncs04}
\bibliography{refs}
\end{document}